\documentstyle[11pt,moriond,epsfig]{article}

\begin{document}

\vspace*{4cm}
\title{CP AND T VIOLATION IN $ K_L \rightarrow \pi^+ \pi^-
e^+ e^- $}

\author{ Bruce Winstein }

\address{The University of Chicago, 5640 South Ellis Avenue, \\
Chicago, IL 60637, U.S.A.}

\maketitle\abstracts{I will here make some brief comments on the
T- and CP-odd asymmetry
observed by KTeV in the $ K_L \rightarrow \pi \pi ee $ decay.}

\section{$K_L \rightarrow \pi^+ \pi^- e^+ e^- $ Studies in KTeV}

The decay $ K_L \rightarrow \pi \pi ee $ stems from a $ \pi \pi \gamma $
decay followed by an internal photon
conversion.  Figure 1 shows the distributions in 
$ \pi \pi $ invarient mass for
$ K_L \rightarrow \pi \pi $ candidates, this from the data set used
for KTeV's first $ \epsilon^{\prime}/\epsilon$ result.\footnote{A.
Alavi-Harati et al., Phys. Rev. Lett. \textbf{83}, 22 (1999).}
There is a low-side tail
in the $ \pi^+ \pi^- $ distribution that
does not show up in the 2$ \pi^0 $ distributions, and that is due to
bremstrahlung.

\begin{figure}[!ht]
\centering
\epsfysize=3.5in   
\hspace*{0in}
\epsffile{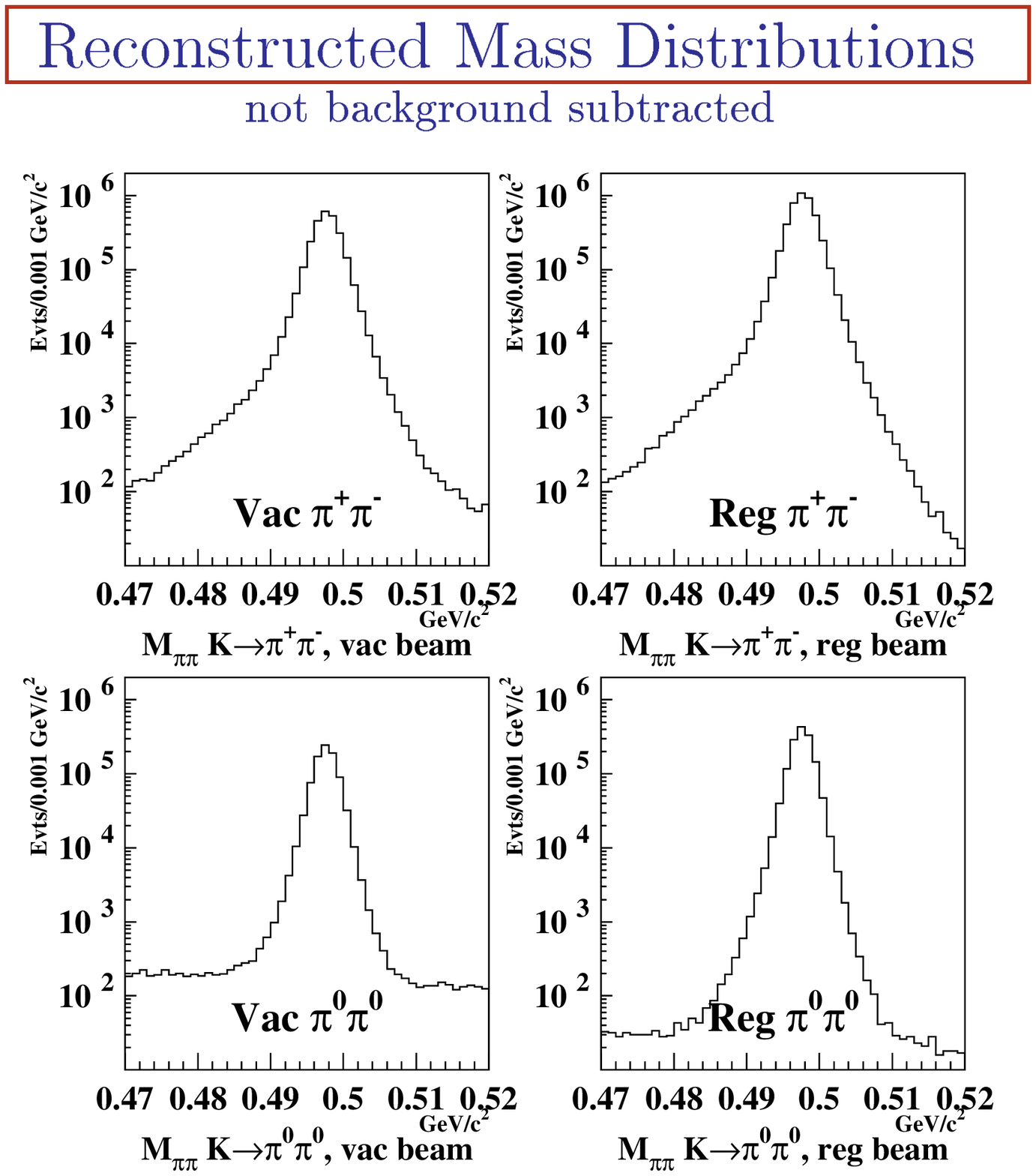}
\caption{$M_{\pi \pi} $ distributions for KTeV data.}
\end{figure}

Another process, that of direct emission, also results in the same 
$ \pi \pi \gamma $
final state.  These two processes cannot directly interfere since
they are of
different multipoles.  However, if the gamma is internally
converted, as
was pointed out by Sehgal and Wanninger,\footnote{L.M. Sehgal
and M. Wanninger, Phys. Rev. \textbf{D46}, 1035 (1992).} interference
is possible.

These two processes have been studied by KTeV and other groups,
and they
are of comparable magnitude.\footnote{E. Ramberg et al., Phys. Rev.
Lett. \textbf{70}, 2525 (1993).}  Since the inner-brem contribution
is CP
violating while the direct emission is CP conserving, one has
a favorable
situation for novel interference effects.

The interference is observable in the $ \Phi $ distribution in the
$ K_L \rightarrow \pi \pi ee $ decay,\footnote{J. Adams et al.,
Phys. Rev. Lett. \textbf{80}, 4123 (1998).}
where $ \Phi $ is the angle between the planes formed by the pions
and by the
electrons.  The predicted distribution is of the form:

\[ dN/d \Phi = A + \textrm{Bsin}^2 (\Phi) + \textrm{Csin}(2 \Phi) \]

The latter term is proportional to:

\[ \textrm{sin} (\Phi)\textrm{cos}(\Phi) = (n_{\pi} \times n_e) \bullet
p_{\pi}/|p_{\pi}| (n_{\pi} \bullet n_e), \]

where $ n_{\pi} $, $ n_e $ are the normal unit vectors to the
$ \pi $, e planes,
respectively, and $ p_{\pi} $ is the momentum of the two pions, all
in the kaon
center-of-mass system.  Since each normal itself is a
cross-product of two
vectors, the above quantity is a product of 9 vectors and is
both CP- and
T-odd.

Figure 2 shows the 4-particle invarient mass distribution in KTeV at the
earliest stages of the analysis, showing a clear peak at the kaon mass.
Also shown is the asymmetry (in the number of events with sin(2$ \Phi $)
greater
than zero vs. those less than zero) vs. invarient mass.  A very clear
effect is seen at the kaon mass, even with no attempt to suppress
background.  Figure 3 shows the same mass distribution after cuts; here
the background is at the 2\% level.

\begin{figure}[!ht]
\centering
\epsfysize=3.5in   
\hspace*{0in}
\epsffile{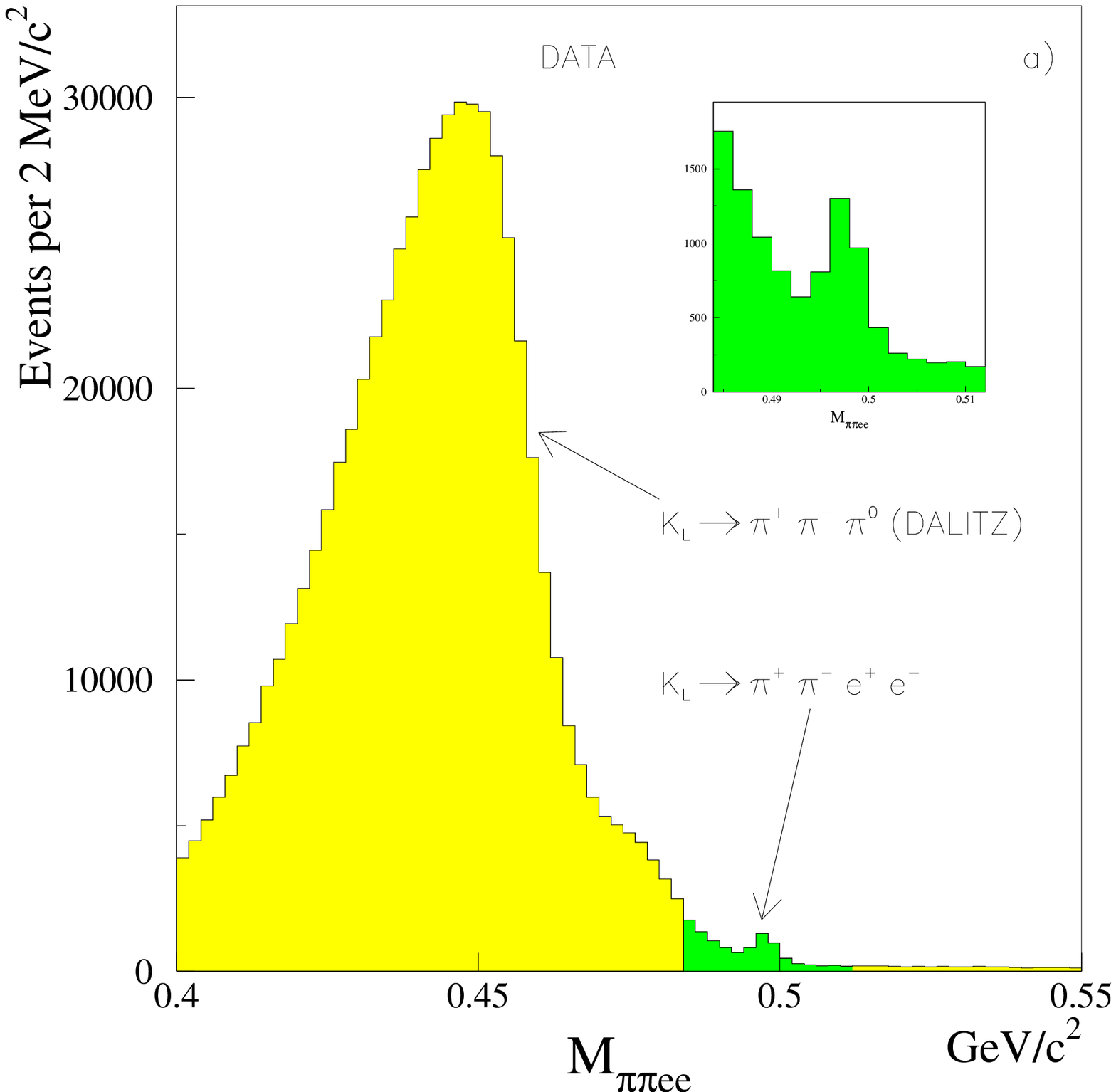}
\end{figure}

\begin{figure}[!ht]
\centering
\epsfysize=3.5in   
\hspace*{0in}
\epsffile{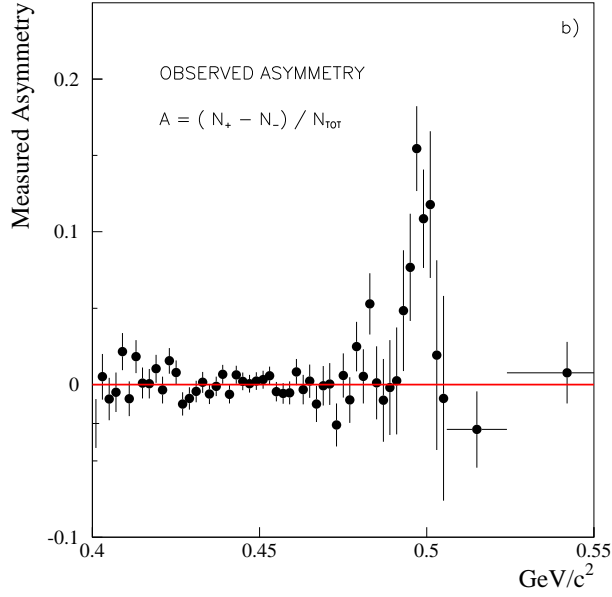}
\caption{The raw $ M_{\pi \pi ee}$ distribution (a); and the
asymmetry vs. $M_{\pi \pi ee} $(b).}
\end{figure}

\begin{figure}[!ht]
\centering
\epsfysize=3.5in   
\hspace*{0in}
\epsffile{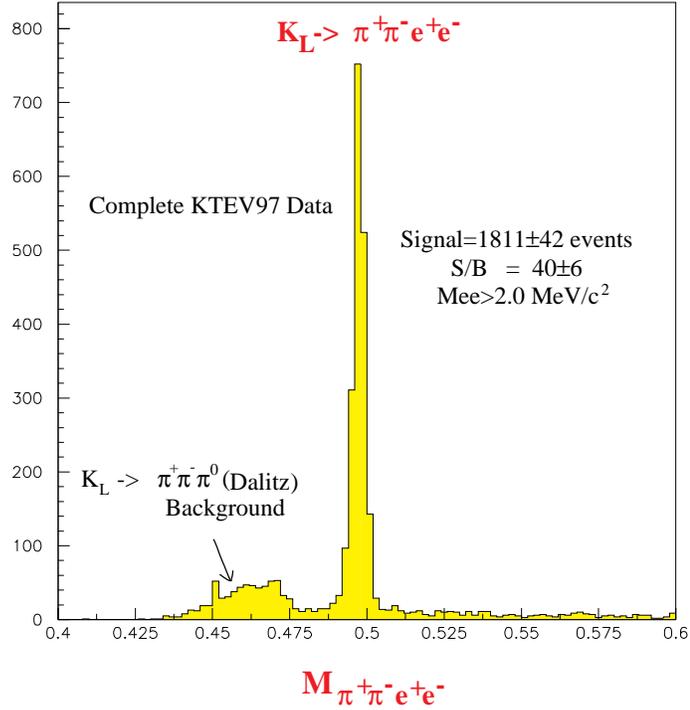}
\caption{$M_{\pi \pi ee} $ distribution after cuts.}
\end{figure} 

We also show in Figure 4 the $ \Phi $ distribution for the
$ \pi^+ \pi^- \pi^0 $ decay with a Dalitz conversion. No asymmetry
is expected; we find (-0.02 $ \pm $
0.05)\%, showing that the detector and analysis procedure do not
produce
an artificial asymmetry.

\begin{figure}[!ht]
\centering
\epsfysize=3.5in   
\hspace*{0in}
\epsffile{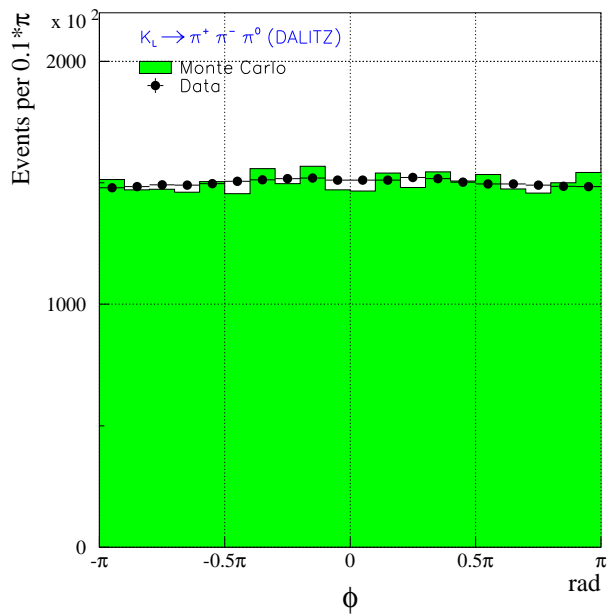}
\caption{Distribution in $ \Phi $ for $K_L \rightarrow
\pi^+ \pi^- \pi^0 $ Dalitz decays, showing no instrumental asymmetries.}
\end{figure} 

Before extracting the final asymmetry, we first need to study the form
factor in the direct decay and then we need to calculate and correct for
the detector acceptance.

The form factor is traditionally described by the following expression:  
\[ F = \tilde{g}M1 \{ 1 + \frac{a_1/a_2}{(M^2_{\rho} - M^2_K + 2 M_K
(E_{e+} + E_{e-})} \}  \]

We have determined the quantity a1/a2 both using the $ \pi \pi ee $
and $ \pi \pi \gamma $ decays,
finding consistent results:\footnote{S. Ledovskoy, contribution to
the Chicago Conference on Kaon Physics, June, 1999, to be published
in the proceedings, U. of Chicago Press, Editors J. Rosner and B. Winstein.}

\[ a1/a2 = -0.72 \pm 0.03 (\textrm{stat}) \pm  0.009
(\textrm{syst})~~~~~~~~~~(\pi \pi ee); \textrm{and}, \]
\[ a1/a2 = -0.729 \pm 0.026 (\textrm{stat}) \pm 0.015
(\textrm{syst})~~~~~~~~~~(\pi \pi \gamma). \]

The expected distributions in $ \Phi $ are shown in Figures 5 and 6
superimposed upon the
data.  There is good agreement, the observed asymmetry being:

\[ A_{\Phi} = 23.3 \pm 2.3 (\textrm{stat})\%~~~~~~~~~~\textrm{(raw asymmetry)}.\]

\begin{figure}[!ht]
\centering
\epsfysize=3.5in   
\hspace*{0in}
\epsffile{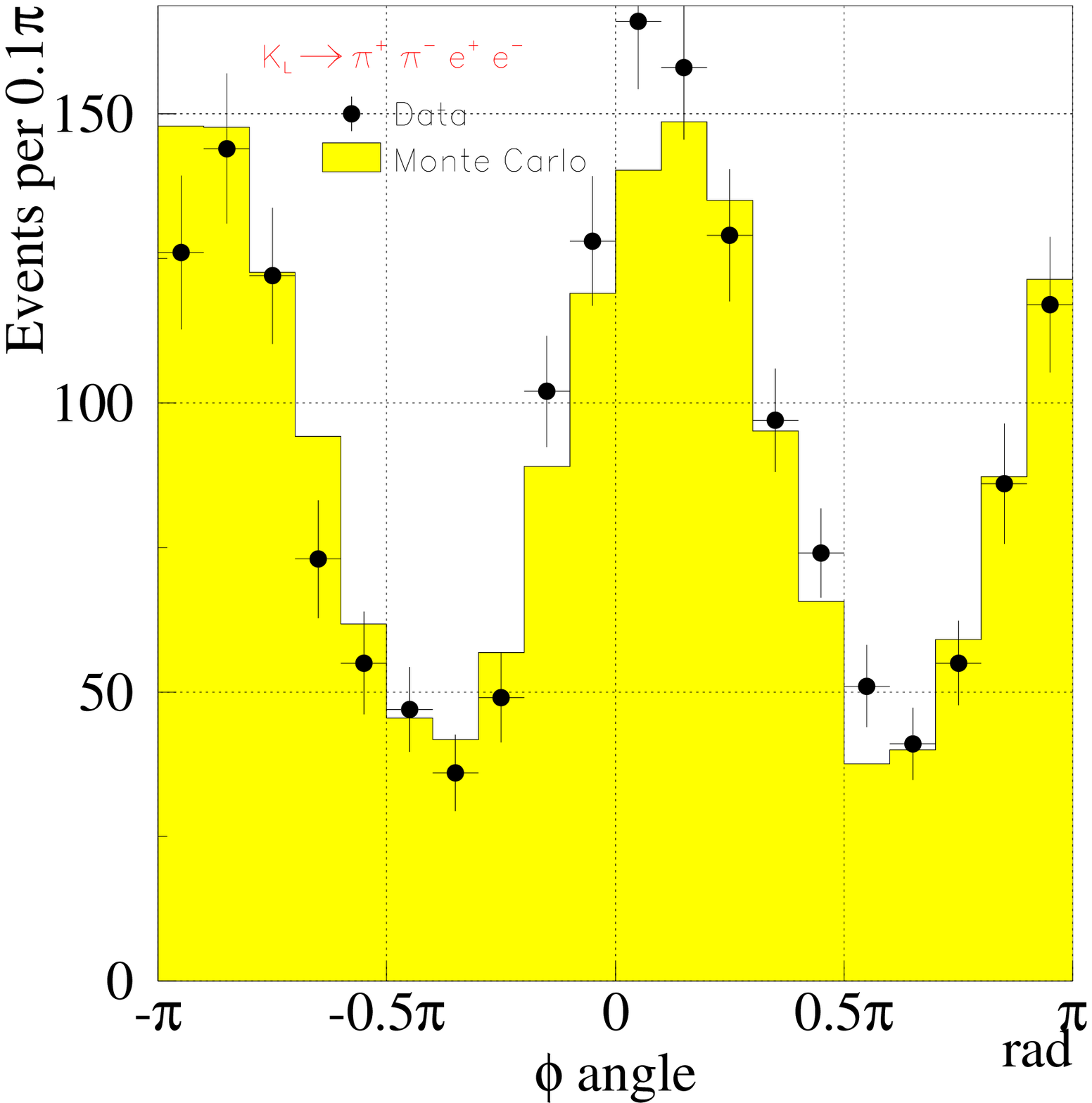}
\caption{Distribution in $ \Phi $ for $K_L
\rightarrow
\pi \pi ee $ decays, data and monte carlo.}
\end{figure}

\begin{figure}[!ht]
\centering
\epsfysize=3.5in   
\hspace*{0in}
\epsffile{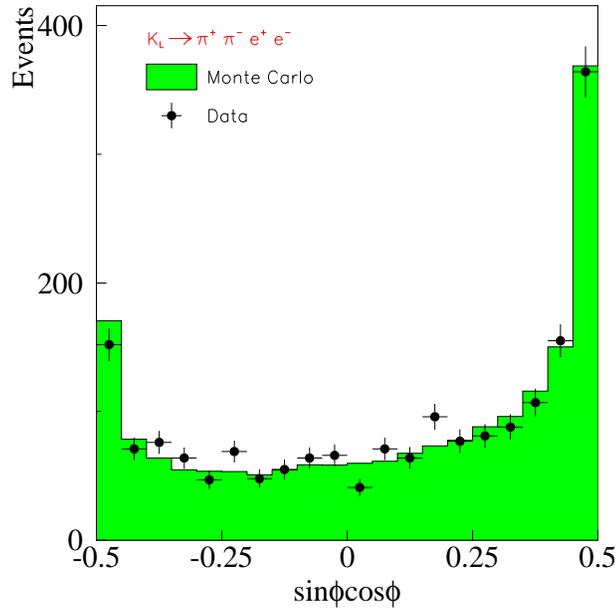}
\caption{Distribution in sin$ \Phi $cos$ \Phi $
for $ K_L \rightarrow \pi \pi ee $
decays, data and monte carlo.}
\end{figure} 

This is certainly the largest CP asymmetry yet observed.  It is an
``$ \epsilon $''
effect, just like the other indirect effects observed to date:
$ K_L \rightarrow \pi \pi $, $ K_L \rightarrow \pi l \nu $,
$ K_{L,S} \rightarrow \pi \pi \gamma $.  It is so large
because of the happy circumstance that the
two interfering amplitudes are of comparable size, allowing an
amplification factor of two orders of magnitude.

The acceptance actually enhances the observed asymmetry.  The interference
is maximal at ee invarient masses around 60 MeV, where we have good
acceptance, and falls to zero as the ee mass goes to zero, a region where
we have poor acceptance.  Hence the acceptance corrected asymmetry is
smaller than the observed asymmetry:

\[ A_{\Phi} = 13.6 \pm 2.5 (\textrm{stat}) \pm 1.2
(\textrm{syst})\%~~~~~~~~~~\textrm{(corrected asymmetry)}. \]

The value we see is in good agreement with that predicted, namely 14.4\%.

\section{Is Time Reserval Symmetry Violated?}
	
We now discuss the issue of whether this observation is manifest
T-violation.

	The effect we see is clearly CP-odd and T-odd.  One issue that can
affect our conclusions is final state interactions (FSI).  The relevant FSIs for
this reaction are strong -- leading to phase shifts, and electromagnetic.
Since both conserve CP, the observed asymmetry is manifest CP
violation.

	The same cannot be said for T-violation.  But we can argue that
strong FSIs cannot \textbf{fake} an effect.  The effect we see comes about because
of the non-zero value for $ \eta_{+-} $ and the strong phase shifts only serve to
moderate the effect, not to generate such an effect.

	What about electromagnetic FSIs?  The only process that can alter
the distribution between the $ \pi $ and e planes is the exchange of photons
between the pions and electrons.  Such diagrams will indeed slightly alter
the distribution in the $ \Phi $ angle, but they will \textbf{not} generate, or even
moderate, an asymmetry.  This can be shown by construction but it also
results from the knowledge that such an asymmetry would be CP violating
and FSIs (where the interaction is CP conserving) cannot generate CP
violating effects.

	So we conclude that FSIs cannot generate the asymmetry we have
observed.  Can we conclude, therefore, that we have unambiguously seen
T-violation?  This subject is under intense discussion in the literature
and we will briefly recap the situation here.

	One issue is whether, from this asymmetry observation alone, one
can conclude T-violation.  Wolfenstein\footnote{L. Wolfenstein, Phys. Rev.
Lett. \textbf{83}, 911-912 (1999).} has pointed out that were the
phase of $ \eta_{+-} $ closer to $ 135^0 $ rather than $ 45^0 $, and were T a good symmetry,
we would observe exactly the same effect but it would arise from CPT
violation in the mixing of the neutral koans.  Thus his point
is that this decay alone is not enough to
establish T-violation.  Of
course in KTeV and in many other experiments, we observe that $ 135^0 $
can be
definitively ruled out so this objection can be met.

	Ellis and Mavromatos\footnote{J. Ellis and N.E.
Mavromatos, hep-ph/9903386.} have hypothesized a very large CPT violating
effect in the MI $ K_L~\rightarrow~\pi \pi \gamma $ decay which could be responsible for the observed
effect, again in a T-invarient world.  This can probably be ruled out
from other information but we are not sure  at this time.

	Finally (for now), Bigi and Sanda\footnote{I.I. Bigi and A.I.
Sanda, hep-ph/9904484 (revised).} have elected to abandon
unitarity and can reproduce the observed effect (as well as the asymmetry
observed by CPLEAR\footnote{CPLEAR collaboration, Phys. Lett. 
\textbf{B444}, 43-51 (1998).}) by an admittedly unnatural but still possible scheme
with large, finely tuned, CPT violation in a number of channels.  Again
this has other experimental consequences which can perhaps be used to rule
it out.

	We remain confident that  even if these ideas are rejected, others
will be proposed to save the notion of symmetry under time reversal.

\end{document}